\documentclass[%
 aip,
 amsmath,amssymb,
 reprint,%
]{revtex4-1}
\usepackage{xcolor}
\usepackage{graphicx}
\usepackage{dcolumn}
\usepackage{bm}

\usepackage[utf8]{inputenc}
\usepackage[T1]{fontenc}
\usepackage{mathptmx}
\usepackage{etoolbox}

\makeatletter
\def\@email#1#2{%
 \endgroup
 \patchcmd{\titleblock@produce}
  {\frontmatter@RRAPformat}
  {\frontmatter@RRAPformat{\produce@RRAP{*#1\href{mailto:#2}{#2}}}\frontmatter@RRAPformat}
  {}{}
}%
\makeatother
\begin{document}


\title{Prediction and mitigation of nonlocal cascading failures using graph neural networks}
\author{Bukyoung Jhun}
\thanks{These two authors contributed equally}
\affiliation{CCSS and CTP, Seoul National University, Seoul 08826, Korea}

\author{Hoyun Choi}
\thanks{These two authors contributed equally}
\affiliation{CCSS and CTP, Seoul National University, Seoul 08826, Korea}

\author{Yongsun Lee}
\affiliation{CCSS and CTP, Seoul National University, Seoul 08826, Korea}
\affiliation{Center for Complex Systems and KI for Grid Modernization, Korea Institute of Energy Technology, Naju, Jeonnam 58217, Korea}
\author{Jongshin Lee}
\affiliation{CCSS and CTP, Seoul National University, Seoul 08826, Korea}
\affiliation{Center for Complex Systems and KI for Grid Modernization, Korea Institute of Energy Technology, Naju, Jeonnam 58217, Korea}
\author{Cook Hyun Kim}
\affiliation{CCSS and CTP, Seoul National University, Seoul 08826, Korea}
\affiliation{Center for Complex Systems and KI for Grid Modernization, Korea Institute of Energy Technology, Naju, Jeonnam 58217, Korea}
\author{B. Kahng}
\email{bkahng@kentech.ac.kr}
\affiliation{Center for Complex Systems and KI for Grid Modernization, Korea Institute of Energy Technology, Naju, Jeonnam 58217, Korea}
\date{\today}

\begin{abstract}
    Cascading failures (CFs) in electrical power grids propagate nonlocally; After a local disturbance, the second failure may be distant. To study the avalanche dynamics and mitigation strategy of nonlocal CFs, numerical simulation is necessary; however, computational complexity is high. Here, we first propose an avalanche centrality (AC) of each node, a measure related to avalanche size, based on the Motter and Lai model. Second, we  train a graph neural network (GNN) with the AC in small networks. Next, the trained GNN predicts the AC ranking in much larger networks and real-world electrical grids. This result can be used effectively for avalanche mitigation. The framework we develop can be implemented in other complex processes that are computationally costly to simulate in large networks.
\end{abstract}

\maketitle

\begin{quotation}
    A small, localized disturbance can result in catastrophic global failure of an entire network. Therefore, cascading failure (CF) in infrastructure networks such as power grids and Internet, has many practical implications. Recently, this subject has become more critical as power-grid systems are decentralized by various energy sources such as wind and solar power. The electric power from these renewable energy sources may not fit those from traditional energy sources, which causes CFs in the electric power grid. This CF propagates nonlocally, which differs from what occurs in local epidemic disease spread. Here, we study the avalanche dynamics and avalanche mitigation strategy of nonlocal CFs. Using a graph neural network (GNN) trained in small networks, we can successfully predict the avalanche centrality ranking of each node in much larger networks and real-world electrical grids. This result enables effective avalanche mitigation.
\end{quotation}

\section{Introduction\label{sec:intro}}

Small, local disturbances in a complex networks can trigger consecutive failures of other nodes in the network. Most failures remain local and do not last for an extended period~\cite{Ewart1978,JieChen2001,Carreras2004a}. However, there is a small probability that they will spread throughout the network, resulting in catastrophic global failure.
This type of failure in infrastructure networks such as the Internet and electrical grids can result in tremendous financial damage and even the loss of human lives.
For example, in the Northeast blackout of 2003, an initial disturbance in Ohio triggered the largest blackout in North American history, which affected more than 50 million people and lasted for up to 15 hours in the US and Canada.

The prediction and control of cascading failures in complex networks is a central topic of research in network science.
Conventional epidemic models~\cite{Pastor-Satorras2015} based on local contact process, propagating in branching process, fail to capture the unique features of CFs that propagate nonlocally~\cite{Pagnier2019,Hines2016}. A pioneering model associated with the nonlocal CFs~\cite{De1985} was proposed to understand the CF dynamics in random resistor networks, and then this subject has been extensively studied~\cite{gilabert1987,kahng1988}.

Most studies of CFs have used numerical simulations and enumeration. In this study, computational complexity is an important issue. The computational complexity of the local CF dynamics is generally somewhat lower than that of the nonlocal dynamics. For instance, for $k$-core percolation~\cite{Lee2016a}, CF occurs at nodes with a degree of less than $k$. Thus, CFs propagate locally. The calculation of a single cascade step has a time complexity of $O(N)$ in a network with $N$ nodes and can be simulated even in large-scale networks. By contrast, the random resistor network model~\cite{De1985}, in which CFs propagate nonlocally, has a time complexity of $O(N^2)$. To overcome this problem, Batrouni et al.~\cite{Batrouni1986} proposed the Fourier acceleration algorithm, which decreased the computation complexity dramatically, to $O(N\log N)$.
In the Motter--Lai (ML) model~\cite{Motter2002}, CF propagates nonlocally, and the time complexity is higher, $O(N^2\log N)$, because the shortest paths must be identified for a single cascade step. Therefore, the ML model has been simulated only for small networks.

The ML model~\cite{Motter2002} was proposed to study data packet transport along the shortest paths between two routers on the Internet. When heavily loaded nodes break down, network traffic is rerouted, causing load redistribution; consequently, CF can occur at nodes distant from the failed nodes. This simple dynamics of the ML model makes it possible to understand the propagation of CF in detail; thus, the model has been widely implemented ~\cite{Motter2004,Hayashi2005,Zhao2004,Zhao2005,Wu2006,Schafer2006,Holme2002a,Lee2005a,Lai2004,Kornbluth2021,Wang2007,Fang2015}.

In this study, we addressed the problem of avalanche mitigation. We define the avalanche centrality (AC) of each node and show that this quantity can be used to effectively suppress avalanches. To reduce the time complexity of the AC calculation, which exceeds $O(N^3\log N)$, we used a graph neural network (GNN) approach.
The GNN is a deep learning model for analyzing graph data; it was introduced by Scarselli et al.~\cite{scarselli2008}.
In particular, we constructed a scalable GNN structure that is independent of network size using an inductive learning scheme that is applicable to various type of networks. The GNN structure was designed to be effective for modeling a large network, even if it is trained using the simulation results for a small network. An avalanche mitigation strategy that uses this methodology and is applicable to large-scale networks, which are impractical to simulate, is proposed.

\begin{figure}
    \includegraphics[width=0.9\columnwidth]{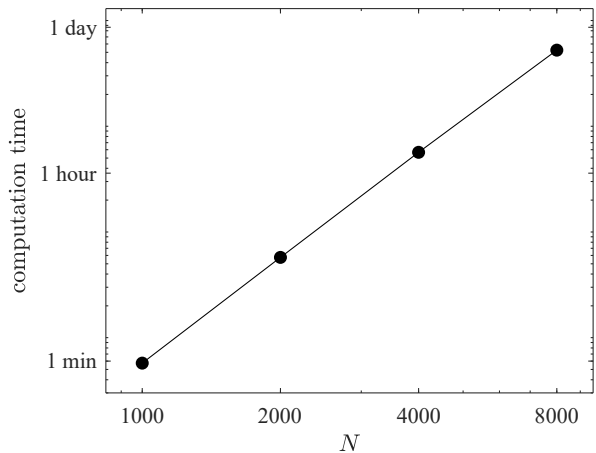}
    \caption{Computation time of ML model as a function of network size. The Schultz--Heitzig--Kurths (SHK) random power grid model was used. The computation time is proportional to $N^{3.3}$; thus, if the network size doubles, the computation time increases by approximately tenfold. More than a year of computation is required to simulate the process in a network of size $N=25000$ using the CPU of model i7-10700.}
    \label{fig:computation_time}
\end{figure}

This paper is organized as follows. First, we introduce the ML model and its underlying network structure in Sec.~\ref{subsec:motter_lai}. In Sec.~\ref{subsec:avalanche_mitigation}, the problem of avalanche mitigation and an effective strategy for modeling the avalanche dynamics are discussed. In Sec.~\ref{subsec:gnn}, the training dataset and performance measure are introduced, along with the scalable GNN structure.
Avalanche mitigation using the GNN is presented for both a large network and a real power grid network.

\begin{figure}
    \includegraphics[width=0.9\columnwidth]{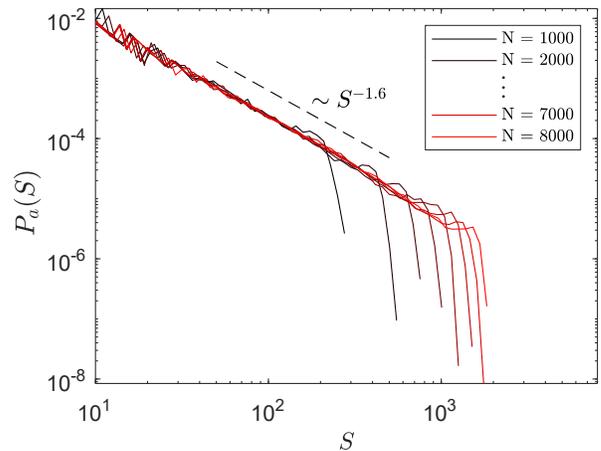}
    \caption{Avalanche size distribution of SHK network with various numbers of nodes for the ML model. The control parameter $a$ is taken as 0.25, at which the avalanche sizes have a scale-free distribution with exponent $\tau \approx 1.6$.}
    \label{fig:avalanche_size_distribution}
\end{figure}

\section{Methods}
\subsection{Avalanche dynamics} \label{subsec:motter_lai}
Many models of cascading failure dynamics have been proposed, for example, random fuse networks~\cite{De1985}, $k$-core percolation~\cite{Goltsev2006,Dorogovtsev2006,Baxter2015,Lee2016a}, dynamically induced cascades~\cite{Schafer2018}, and others~\cite{Crucitti2004,Kinney2005,Buldyrev2010,Cellai2013,Baxter2012,Reis2014,Lee2016b}.
Among them, the ML model~\cite{Motter2002} is one of the simplest models capable of capturing the avalanche dynamics of propagation through nonlocal nodes. It has been applied to power grids~\cite{Kornbluth2021}.
This simple, shortest-path-based model can be used to understand the avalanche process.
Consequently, it has been extensively applied in studies of suppression strategies~\cite{Motter2004,Hayashi2005}, the analytic calculation of the robustness criterion~\cite{Zhao2004,Zhao2005}, the effect of network topology~\cite{Wu2006}, and other topics~\cite{Schafer2006,Holme2002a,Lee2005a,Lai2004,Kornbluth2021,Wang2007,Fang2015}.

In the ML model~\cite{Motter2002}, the capacity $C_i$ of each node $i$ is proportional to the betweenness centrality (BC), $B_i^{(0)}$, of node $i$, as follows:
\begin{equation}
    C_i = (1+a) B_i^{(0)} \,,
\end{equation}
where $a>0$ is a model parameter. The superscript (0) denotes that the BC is calculated in a network without any failure.
The BC is the number of shortest paths that pass through the node; it is written as
\begin{equation}
    B_i^{(0)} = \sum_{j\neq k \neq i} \frac{m_{jk}(i)}{m_{jk}} \,,
\end{equation}
where $m_{jk}$ is the number of shortest paths between nodes $j$ and $k$, and $m_{jk}(i)$ is the number of those paths that pass through node $i$.

We initiate the avalanche dynamics by causing node $i$ on the network to fail. The failed node is effectively removed from the network, as all the connected links are disconnected.
Then, all the shortest paths that had passed through node $i$ are rerouted to detours. Consequently, the excess BC is redistributed over the network.
The BC of each node $j$ is then updated to $B_j^{(1)}$.
If the updated $B_j^{(1)}$ exceeds the capacity of the preassigned node, $C_j$, node $j$ is overloaded and fails. Other nodes with BCs exceeding their capacities fail simultaneously. Then the BCs of the remaining active nodes are again updated to $B_k^{(2)}$. The process is repeated until there are no overloaded nodes in the network: $B_k^{(t)} < C_k$ for all remaining nodes $k$.

The avalanche size $S_i$ of each node ($i=1,\cdots, N$) has a distribution $P_a(S)$. On scale-free networks, the distribution exhibits power-law behavior as $P_a(S)\sim S^{-\tau}$, where $\tau\approx 2.1$ when a particular value of $a$ is chosen, which is denoted as $a_c\approx 0.15$~\cite{Lee2005a}. This result may indicate that $P_a(s)$ exhibits critical behavior at $a_c$. The exponent $\tau$ is insensitive to the degree exponent $\lambda$ of scale-free networks and seems to be closely related to the exponent of the diameter-change distribution when nodes are deleted one by one~\cite{Kim2003}. It has also been found that the average size $\langle s (k) \rangle$ of avalanches triggered by removing each node with degree $k$ depends on $k$ as $\langle s(k)\rangle \sim k^{(\lambda-1)/(\tau-1)}$. Note that the avalanche size distribution observed in cascading failures in real-world electric power grids has a heavy-tailed distribution ~\cite{Carreras2016,Hines2009,Dobson2007,Carreras2004,Carreras2004a}.



In this study, we simulated the ML model on a network model proposed by Schultz, Heitzig, and Kurths (SHK), which was designed according to the essential features of real-world electric power grids~\cite{Schultz2014}. This model has many control parameters. Using an appropriate set of parameter values used in Ref.~\cite{Nitzbon2017}, we obtained power-law behavior of the avalanche size distribution at $a_c \approx 0.25$. The exponent $\tau\approx 1.6$ was found as shown in Fig.~\ref{fig:avalanche_size_distribution}, which is close to $\tau=1.5$, the value of the sandpile model in random networks~\cite{Goh2003}.

The ML model was modified for application to electric power grids by Carreras et al.~\cite{Carreras2002}. In this modified ML model, each node $i$ is characterized by an input power $P_i$, which has positive and negative values for power generators and consumers, respectively. To account for the avalanche failures of links rather than nodes, the capacity of each link is defined as $C_{ij}=(1-a)F_{ij}^{\rm max}$, where $F_{ij}^{\rm max}$ is the maximum power flow through link $(ij)$ between nodes $i$ and $j$. At each time step, the direct-current circuit equation is solved, and the power flow $F_{ij}$ through link $(ij)$ is obtained. If $F_{ij} > C_{ij}$, link $(ij)$ is disconnected, and the power flow of each link is recalculated. This link disconnection process is repeated until further link failure occurs. The avalanche size is obtained by counting the number of failed links.

\subsection{Avalanche mitigation strategy}
\label{subsec:avalanche_mitigation}

\begin{figure*}
    \includegraphics[width=\textwidth]{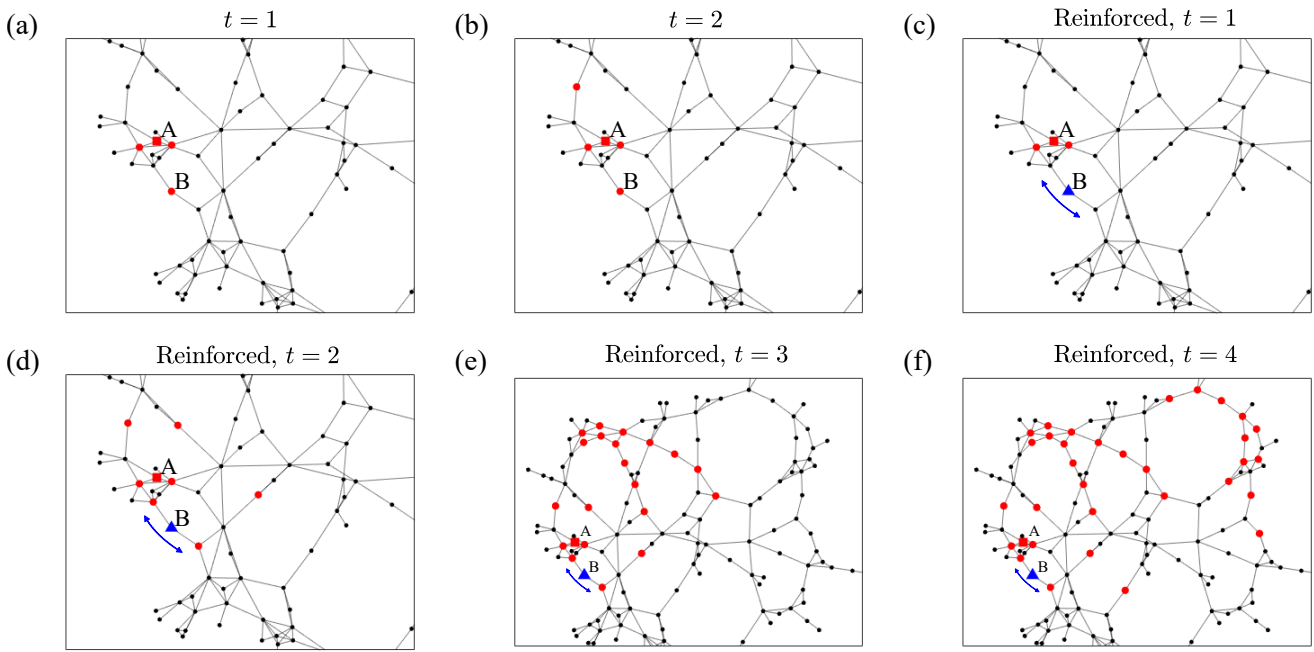}
    \caption{Paradoxical effect of reinforcement in electric power grid of France.
        Red square (indicated by A) represents the initial failure, red dots represent secondary failures, and blue triangle (indicated by B) represents a reinforced node. (a) and (b) If no node is reinforced, the avalanche ends at $t=2$ with avalanche size $S=5$. (c)--(f) If node B (\textcolor{blue}{$\blacktriangle$}) is reinforced, the avalanche lasts until $t=4$, and the avalanche size increases to 33. Because the reinforced node does not fail, heavy traffic through route via node B (blue arrow) is maintained, causing other nodes on the route to fail. Reinforcement thus may result in a significantly larger global cascade.}
    \label{fig:vaccination_paradox}
\end{figure*}

Various methods of reducing the avalanche size in the ML model have been implemented. For instance, after an initial failure but before the failure propagates to other nodes, selected nodes are removed to reduce the avalanche size~\cite{Motter2004,Hayashi2005}. However, the initial failure often propagates so rapidly that it is impossible to remove nodes involved in significant propagation of the failure. As another strategy, a parameter $a$ is assigned to each node under the constraint $\sum_i a_i=\mathrm{const}$~\cite{Wang2007}. However, there is no way to adjust the capacity of every node to an appropriate value to minimize the avalanche size.

In this study, we compare the performance of various strategies that minimize the avalanche size when nodes are reinforced. To simplify the problem, we suppose that reinforced nodes never fail unless they are intentionally chosen as the initial failure of the cascade; however, the effectiveness of the strategies should be maintained even when reinforcement is finite.

The problem with reinforcement is analogous to the problem of vaccination against epidemic contagion, where epidemic spreading is minimized by vaccination; however, there is a crucial difference. Specifically, during epidemics, a node is eliminated after vaccination, and the centrality measures of the network are recalculated to determine the next node to be vaccinated ~\cite{Wang2017,Clusella2016,Altarelli2014,Cohen2003,Matamalas2018}. This process is repeated until the feasible number of vaccinations is reached. In the avalanche mitigation problem, however, reinforced nodes can remain in the power flow; they simply do not fail even if the power flow increases.

Avalanche mitigation poses a novel type of vaccination problem in which the effects of reinforced nodes on the avalanche are complex. A unique feature of the avalanche mitigation problem is the paradoxical effect of reinforcement. In some cases, reinforcement of a node can increase the avalanche size. As shown in Fig.~\ref{fig:vaccination_paradox}, even if one correctly predicts that node B is at risk and reinforces it, reinforcement results in a significant increase in avalanche size. Therefore, reinforcement of nodes without a systematic strategy can be counterproductive.

To characterize the role of each node in the avalanche dynamics, we define a binary variable $x^i_j$, which is 1 if node $j$ fails in an avalanche triggered by node $i$ and is 0 otherwise ($x^i_i = 1$). Then, the avalanche fraction of node $i$ is given by $s_i=\sum_j x^i_j/N$. This quantity represents the effect of node $i$ in cascading failures. The failure fraction $f_j$ of node $j$ is defined as $f_j=\sum_i x^i_j/N$ and represents the probability that node $j$ fails in isolation or because of triggering by another node. These two quantities characterize the role of each node in cascading failures. For instance, if $s_i \gg 0$ but $f_i \ll 1$, then node $i$ triggers a large avalanche, but its effect on the avalanche is rather limited.

\begin{figure}
    \includegraphics[width=\columnwidth]{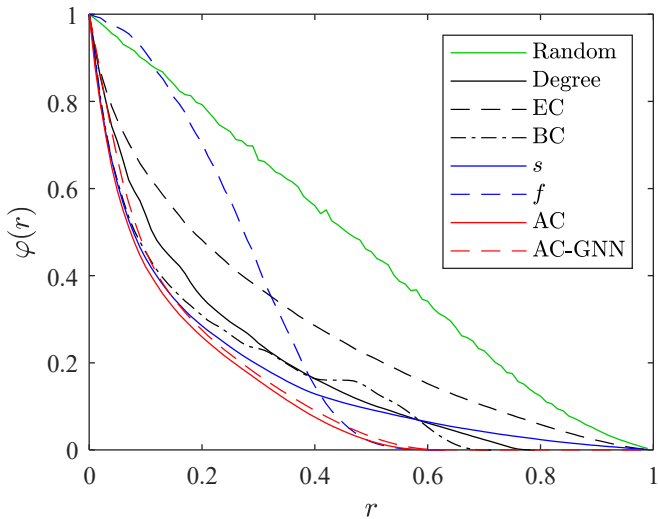}
    \caption{Performance of avalanche mitigation strategies in the SHK network of size $N=1000$. Degree, eigenvector centrality (EC), betweenness centrality (BC), avalanche fraction ($s$), failure fraction ($f$), avalanche centrality (AC), and GNN estimation of avalanche centrality (AC-GNN) are employed for avalanche mitigation strategies. Nodes are reinforced in descending order of the corresponding centrality measure. The area under each curve is the performance measure $R_m$.}
    \label{fig:performance}
\end{figure}

We define the AC of node $i$ as
\begin{align}
    A_i = s_i \left(f_i - \frac{1}{N}\right) \,,
\end{align}
where the factor $1/N$ accounts for the case where node $i$ is selected as the initial trigger of the avalanche. Fig.~\ref{fig:performance} shows that the avalanche size can be reduced by reinforcing nodes with high ACs.

We define the performance measure $R_{m}$ of each avalanche mitigation strategy as
\begin{equation}
    R_m= \int_0^1 \varphi(r) dr \simeq \sum_i \frac{\varphi(r_i)+\varphi(r_{i+1})}{2} \Delta r \,,
\end{equation}
where
\begin{equation}
    \varphi(r) = \frac{\bar{s}(r) - 1/N}{\bar{s}(0) - 1/N} \,.
\end{equation}
Here $r$ denotes the fraction of reinforced nodes, and $\bar{s}(r)$ is the mean avalanche fraction; that is, $\sum_i s_i(r)/N$ when a fraction $r$ of nodes are reinforced according to the avalanche mitigation strategy. The term $1/N$ is needed to exclude initial failure from $s$.
Consequently, $\varphi(0)=1$, $\varphi(1)=0$, and $0\leq R_m \leq 1$. As $R_m$ decreases, the avalanche mitigation strategy becomes more efficient. To calculate $R_m$, we use $\Delta r=0.01$ for small networks ($N\leq 1000$) and $\Delta r=0.1$ for large networks ($N>1000$) to avoid excessive computational cost.

Fig.~\ref{fig:performance} shows the $\varphi(r)$ values of various avalanche mitigation strategies in the SHK network.
The area under each curve represents the performance measure $R_m$.
Reinforcing nodes with high network centralities, such as degree centrality, eigenvector centrality, and BC, blocks the avalanche dynamics more effectively than random reinforcement.
The strategy based on an avalanche fraction $s$ is effective in a small $r$ range. The strategy based on the failure fraction $f$ is even less effective than random reinforcement for small $r$. However, for sufficiently large $r$, strategies based on $s$ and $f$ both become more effective than other strategies except that based on the AC, and the cascading failure is reduced to zero at a certain value $r_c$. This value corresponds to the threshold of herd immunity in epidemics.
The strategy based on the AC that takes into account both the avalanche and failure fractions becomes the most effective strategy among all the strategies in the broad range $r < r_c$.

Table~\ref{tab:performance} lists the performance measure $R_m$ for various avalanche mitigation strategies in the SHK networks of various sizes and real-world electrical power grids. The AC-based strategy exhibits fairly good performance compared with all other methods. We remark that the BC-based strategy is also highly effective. However, it is not scalable because of the logarithmic correction of the computational complexity $O(N^2\log N)$. Thus, strategies using scalable centrality measures such as degree centrality and eigenvector centrality can be employed in the GNN.

\begin{table*}
    \caption{\label{tab:performance}
        Performance measure $R_m$ of mitigation strategies: random reinforcement, reinforcement in descending order of degree, eigenvector centrality (EC), betweenness centrality (BC), avalanche size, failure fraction, avalanche centrality (AC), and GNN prediction of AC. Smaller values indicate better mitigation. The most efficient mitigation strategy is marked by bold.}
    \begin{ruledtabular}
        \begin{tabular}{lcccccccc}
            Network        & Random & Degree & EC     & BC     & Avalanche fraction & Failure fraction & AC              & AC (GNN-pred)
            \\
            \colrule
            SHK ($N=1000$) & 0.4593 & 0.1923 & 0.2767 & 0.1763 & 0.1691             & 0.2697           & \textbf{0.1338} & 0.1467        \\
            SHK ($N=2000$) & 0.4547 & 0.1733 & 0.2582 & 0.1624 & 0.1560             & 0.3102           & \textbf{0.1273} & 0.1377        \\
            SHK ($N=4000$) & 0.4490 & 0.1508 & 0.2423 & 0.1415 & 0.1360             & 0.3569           & \textbf{0.1126} & 0.1234        \\
            SHK ($N=8000$) & 0.4458 & 0.1322 & 0.2199 & 0.1248 & 0.1202             & 0.4431           & \textbf{0.1009} & 0.1107        \\
            Spain          & 0.4887 & 0.3347 & 0.3604 & 0.2648 & 0.2815             & 0.2410           & \textbf{0.2320} & 0.2347        \\
            France         & 0.4855 & 0.3727 & 0.4593 & 0.2477 & 0.2633             & \textbf{0.2171}  & 0.2177          & 0.2306        \\
            UK             & 0.4619 & 0.4220 & 0.2720 & 0.3607 & 0.3203             & \textbf{0.2715}  & 0.2787          & 0.4227        \\
        \end{tabular}
    \end{ruledtabular}
\end{table*}

\subsection{Graph neural network (GNN)} \label{subsec:gnn}
The AC can be used for effective avalanche suppression; however, it has high computational complexity as high as $O(N^3\log N)$. Thus, it cannot be calculated directly in large networks. To overcome this problem, we apply the GNN, a deep learning algorithm that can be used for graph-structured data~\cite{scarselli2008}. We constructed a GNN structure applicable to networks of different sizes. This GNN was trained using the ACs obtained by simulations of small networks. Then, the AC of each node in large networks, where simulations using the ML model are not feasible, can be predicted.

We used the SHK network, a synthetic network mimicking a power grid, to simulate avalanche failures. The network size was selected at random from 100 to 999 in uniform increments. The dataset consists of $10^4$ data.

\begin{figure}[!t]
    \includegraphics[width=\columnwidth]{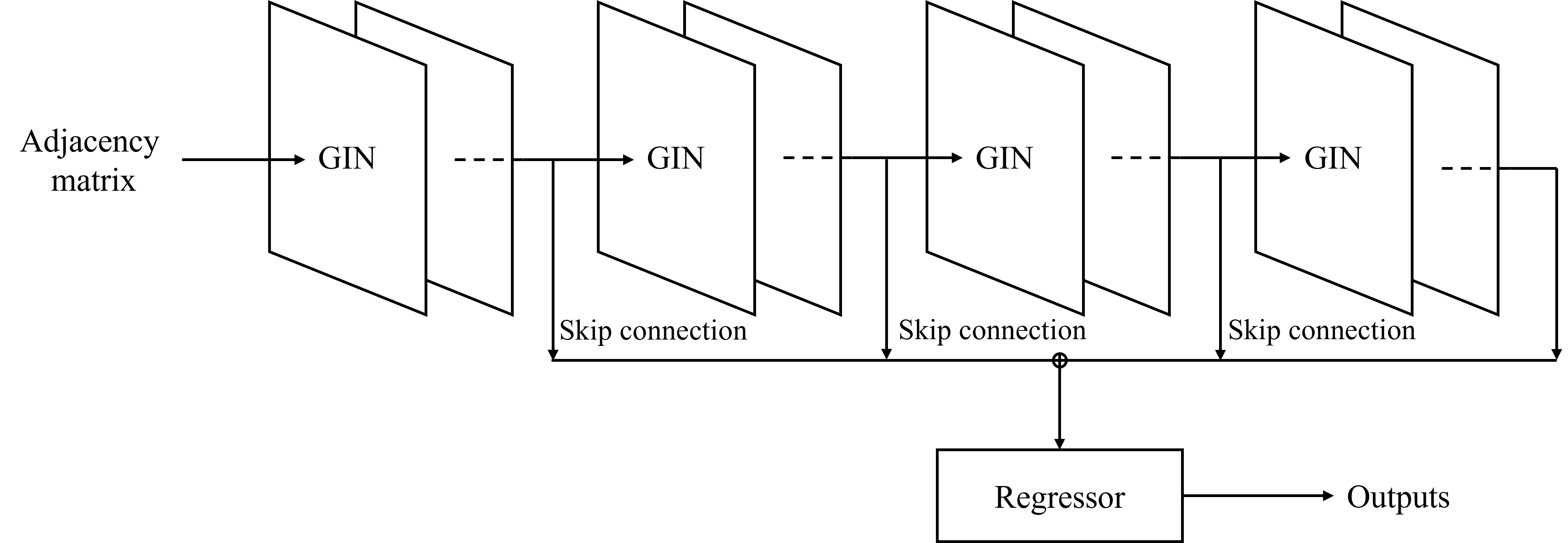}
    \caption{Structure of GNN. Only the adjacency matrix of the network is input, and the features of all nodes are initially constant. Batch normalization and ReLU activation are applied between each pair of adjacent GIN layers, but they are omitted from the diagram for simplicity. A short skip connection is applied between each pair of adjacent GIN layers, and an elementwise average is obtained before the regressor. The regressor is implemented as a single--layer perceptron followed by a sigmoid function for normalization.}
    \label{fig:gnn_structure}
\end{figure}

The GNN is a type of space-based convolutional GNN\cite{Wu2020} composed of graph isomorphism network (GIN) layers~\cite{Xu2018}, as shown in Fig.~\ref{fig:gnn_structure}.
In the GIN layers, the hidden feature $y_i$ of node $i$ is updated according to its own value and those of the nearest--neighbor nodes of $i$, that is, $\{y_j\}$, where $j\in$ n.n. of $i$ as
\begin{equation} \label{eq:gin}
    y_i^\prime = h_\Theta\left((1+\epsilon)y_i + \sum_{j~\in~{\rm{n.n.~of~}} i} y_j\right),
\end{equation}
where $\epsilon$ is a constant and $h_\Theta$ is taken to be a two-layer perceptron for the trainable parameter $\Theta$. The GNN also has a batch normalization layer~\cite{Ioffe2015} and a rectified linear unit (ReLU) activation function between subsequently GIN layers.

The ML model is based on the BC, which contains a global connection information of network. Therefore, the GNN must consider not only local information around a given node, but also the global information of the entire network. As shown in Eq~\eqref{eq:gin}, the GIN layer is used for one-hop calculation. $y_i$ is updated locally. To address the global features at longer hopping distances, many layers should be stacked. This deep structure may cause the so-called vanishing gradient problem~\cite{He2016}. To mitigate this problem and to take into account the effect of nodes at various hopping distances~\cite{Abu2019, Veit2016}, the short skip connection algorithm~\cite{Simonyan2014} is applied for every pair of adjacent layers. Table.~\ref{tab:hp} lists the hyperparameters used to train the GNN.

\begin{table*}[!t]
    \caption{\label{tab:hp} Hyper parameters used to train GNN.}
    \begin{ruledtabular}
        \begin{tabular}{lcccccc}
            name  & node embedding dimension & batch normalization momentum & optimizer & learning rate & loss & L2 regularization \\
            value & 128                      & 0.1                          & RMSprop   & $10^{-3}$     & MAE  & $10^{-5}$
        \end{tabular}
    \end{ruledtabular}
\end{table*}

To perform avalanche mitigation in the GNN framework, it is necessary to rank the ACs of the nodes. This information makes it possible to handle all reinforcement states because one can choose a node set depending on the available reinforcement resources. Because the AC distribution is highly skewed, simple node feature regression is not an appropriate approach. Instead, we applied the quantile transformation to the ACs and then the min-max scale to the obtained values for normalization. The validation and test datasets were scaled using the scaler fitted to the training dataset.

Let $z_i$ is an observable, for example, the AC of node $i$. Rank($i$) is defined according to the relative size of $z_i$ among others. $\sigma(i)$ is the estimated rank of $z_i$ using the GNN. A pair of observables $\left(z_i, z_j\right)$ is  concordant if the order of their ranks is correctly estimated: for the $\mathrm{Rank}(i) < \mathrm{Rank}(j)$, $\sigma(i) < \sigma(j)$. The pair is discordant, otherwise.

A well-known metric for ordinal association is the Kendall rank correlation or Kendall's tau~\cite{Kendall1938}, which is defined as
\begin{equation}\label{eq:kt}
    \tau = \frac{n_c - n_d}{n_c + n_d},
\end{equation}
where $n_c$ and $n_d$ denote the numbers of concordant and discordant pairs, respectively. This Kendall's tau is expressed explicitly as
\begin{equation}
    \tau = \frac{2}{N(N-1)} \sum_{i<j} \mathrm{sign}\left(z_i-z_j\right) \mathrm{sign}\left(\sigma(j)-\sigma(i)\right) \,.
\end{equation}
In the calculation of Kendall's tau, the incorrect ranking of two values that differ greatly receives the same penalty as the incorrect ranking of two similar values. However, the former error is likely to be disproportionately detrimental (in avalanche mitigation, for example). This problem may be more important for highly heterogeneous data.

\begin{figure}[!t]
    \includegraphics[width=0.9\columnwidth]{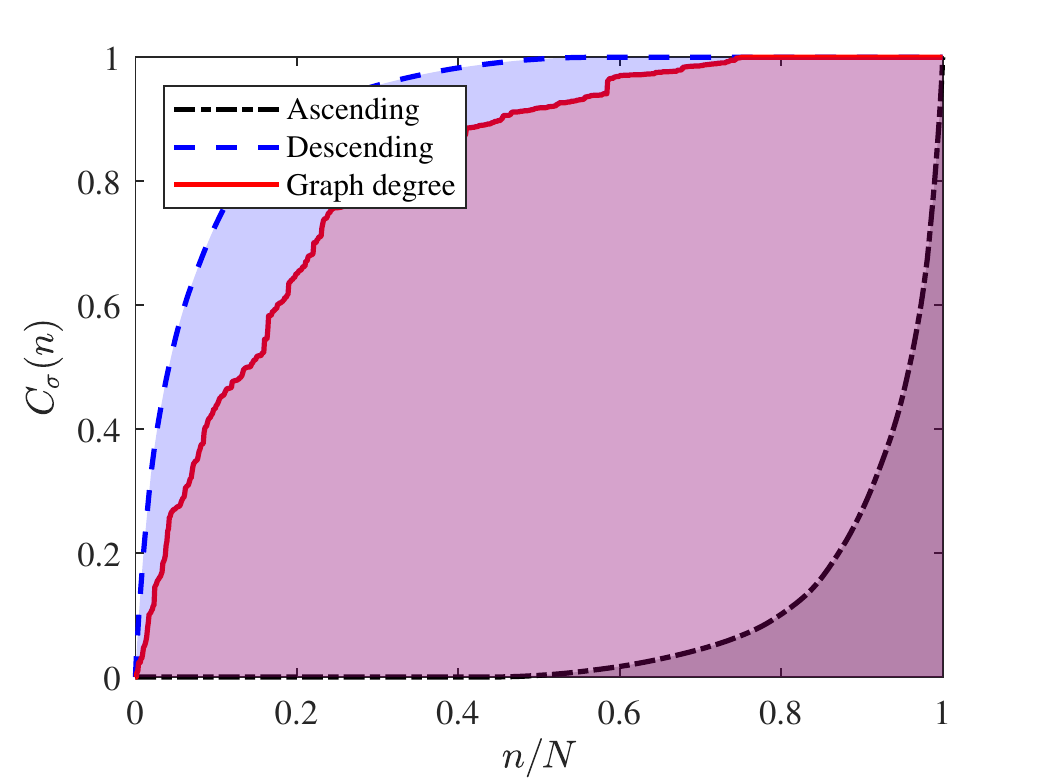}
    \caption{The cumulative fraction $C_\sigma(n)$ of AC as a function of $n$ on an SHK network of size $N=1000$. The mean cumulative fraction $\langle C_\sigma \rangle$ is the area under the cumulative fraction $C_\sigma(n)$ curve. A large $\langle C_\sigma \rangle$ means that the ranking estimation performs well. Red solid curve is obtained by taking the estimated rank $\sigma$ as the graph degree, i.e., the number of connections, of each node. The estimated curve lies between those of the descending order (best case) and ascending order (worst case).}
    \label{fig:mcf_diagram}
\end{figure}

Accordingly, we propose a new quantity, the cumulative fraction $C_\sigma(n)\equiv \sum_{i=1}^{n} z_{\sigma(i)} / \sum_{j=1}^{N} z_j$. This represents the contribution of $n$ the most highly estimated $z_i$s in the sum of all values in the data. The average of $C_\sigma(n)$ over $n$ (denoted as $\langle C_\sigma \rangle$) represents the overall performance of the $\sigma$ estimation.
\begin{align}
    \langle C_\sigma \rangle & = \sum_{n=1}^{N} \frac{1}{N} \sum_{i=1}^{n} \frac{z_{\sigma(i)}}{\sum_{j=1}^{N} z_j} \nonumber \\
                             & = \sum_{i=1}^{N} \frac{N-i+1}{N} \frac{z_{\sigma(i)}}{\sum_{j=1}^{N} z_j} \,.
\end{align}
Fig.~\ref{fig:mcf_diagram} shows the behavior of $C_\sigma (n)$ as a function of $n$. $C_\sigma (n)$ is always between the two extremes: the descending and the ascending orders of Rank($i$) $(i=1,\cdots, N)$. Accordingly, we define the normalized $\langle C_\sigma \rangle$ (denoted as $\langle C_{\sigma,N} \rangle$) as
\begin{equation}
    \langle C_{\sigma,N} \rangle \equiv \frac{\langle C_{\sigma,N} \rangle - \langle C_\mathrm{ascending} \rangle} {\langle C_\mathrm{descending} \rangle - \langle C_\mathrm{ascending} \rangle}.
\end{equation}
Then, $0 \le \langle C_{\sigma,N} \rangle \le 1$.

We also use $R^2$ score to evaluate the prediction of $z_i$ value, which is defined as follows~\cite{r2_score}:
\begin{equation}
    R^2 \equiv 1-\frac{\sum_{i=1}^{N} (z_i-\hat{z}_i)^2}{\sum_{i=1}^{N} (z_i-\bar{z})^2},
\end{equation}
where $\hat{z}_i$ is the estimated value of $z_i$ and $\bar{z}$ is the average of all $z_i$.

\begin{figure*}[!htb]
    \includegraphics[width=0.90\linewidth]{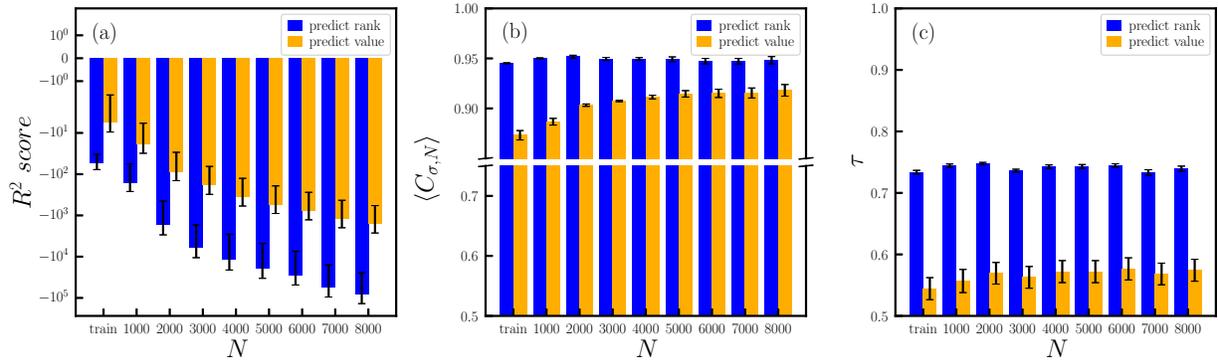}
    \caption{Performance of the GNN prediction in large networks in the range $N=1000-8000$, which is trained in various network sizes from 100 to 999. All the results are averaged over 4 different GNNs with different random seed. (a) $R^2$ scores for different system sizes: the $R^2$ score is not appropriate to predict the performance because the AC values are distributed very skewed. (b) and (c) $\langle C_{\sigma,N} \rangle$ and Kendall's tau for different sizes, respectively. (Blue, dark bars) the GNN predicts ranks of ACs using a quantile scaled dataset. (Orange, light bars) the GNN predicts values of ACs using a raw dataset. Rank prediction using the quantile scaling marked in blue is better and more consistent than value prediction marked in orange. The representation by $\langle C_{\sigma,N} \rangle$ in (b) is more appropriate compared with that by the Kendall's Tau in (c).}
    \label{fig:scalable}
\end{figure*}

Fig.~\ref{fig:scalable} shows the performances of the $R^2$ score, $\langle C_{\sigma,N} \rangle$, and Kendall's tau for the prediction of $z_i$ obtained by the GNN. Although the $R^2$ score decreases rapidly as the network size increases ($N=1000-8000$), the $\langle C_{\sigma,N} \rangle$ and Kendall's tau values of the GNN do not change significantly as $N$ increases. This result suggests that $R^2$ score fails to predict the AC values of networks larger than those in the training dataset. As shown in Fig.~\ref{fig:performance} and Table.~\ref{tab:performance}, node reinforcement according to the AC predicted by the GNN is also effective for suppressing avalanches in the SHK model as well as in several real-world electric power grids of Spain, France, and UK. We remark that the GNN was trained on the SHK model, however, the real-world electrical power grids were not trained.


\section{Conclusion}
In summary, we introduced the concept of the avalanche centrality (AC) of each node in networks and showed that reinforcing nodes in descending order of AC is effective for suppressing the nonlocal avalanche propagation in electrical power grids. However, the calculation of the AC has a high computational cost. Therefore, we employed a GNN to address this problem. We trained the GNN with the ACs of small networks and showed that it can be used for larger networks where the direct calculation of the ACs is not feasible. The GNN predicts the descending order of AC in large networks, allowing the effective suppression of avalanche failures in electrical power grids. Conventionally, the Kendall's tau was used as a performance measure of ranking estimation. However, this measure contains intrinsic drawback: the incorrect estimation of the order of two values that differ greatly is penalized similarly to that of the ranking of two slightly different values. To overcome such a limitation, here, we introduced a new performance measure denoted as $\langle C_\sigma \rangle$ for ranking prediction. This new performance measure successfully reflects the importance of the ordering of two values that differ greatly as shown in Fig~\ref{fig:mcf_diagram}.

In network epidemiology, the microscopic Markov chain approximation (MMCA)~\cite{Gomez2010,Gomez2011,Matamalas2020} was introduced to bypass the computationally cumbersome Monte Carlo simulation.
The result of MMCA was then used to calculate epidemic prevalence or to formulate vaccination strategies~\cite{Matamalas2018,Jhun2021,Jhun2021b}. This method, however, cannot be used to approximate avalanche dynamics that propagates nonlocally, for example, in the ML model.
Therefore, we think the GNN introduced here may replace the role of the MMCA in the analysis of nonlocal avalanche dynamics and the development of mitigation strategy, for example, for the blackout in electrical power grid.

\section*{Data and code availability}
The data and code that support the findings of this study are openly available at https://github.com/CNRC-NAJU/Prediction-and-mitigation-of-nonlocal-cascading-failures-using-graph-neural-networks

\begin{acknowledgments}
    This research was supported by the NRF, Grant No.~NRF-2014R1A3A2069005 and KENTECH Research Grant No. KRG2021-01-007 (BK).
\end{acknowledgments}

%

\end{document}